\documentclass[twocolumn]{aastex61}
\usepackage[figure,figure*]{hypcap}
\usepackage{subfigure}
\usepackage{amsmath}

\shorttitle{A New Standard}
\shortauthors{Jensen-Clem et al.}

\begin{document}

\title{A New Standard for Assessing the Performance of High Contrast Imaging Systems}

\author{Rebecca Jensen-Clem}\altaffiliation{Miller Fellow} \affiliation{Astronomy Department, University of California, Berkeley, Berkeley, CA 94720, USA}

\author{Dimitri Mawet}\affiliation{Department of Astrophysics, California Institute of Technology, 1200 E. California Blvd., Pasadena, CA 91101, USA}
\author{Carlos A. Gomez Gonzalez} \affiliation{Space sciences, Technologies and Astrophysics Research (STAR) Institute, Universit\'{e} de Li\`{e}ge, 19 All\'{e}e du Six Ao\^{u}t, B-4000 Li\`{e}ge, Belgium}\affiliation{Universit\'{e} Grenoble Alpes, IPAG, F-38000 Grenoble, France}
\author{Olivier Absil}\altaffiliation{F.R.S.-FNRS Research Associate} \affiliation{Space sciences, Technologies and Astrophysics Research (STAR) Institute, Universit\'{e} de Li\`{e}ge, 19 All\'{e}e du Six Ao\^{u}t, B-4000 Li\`{e}ge, Belgium}
\author{Ruslan Belikov} \affiliation{NASA Ames Research Center, Moffett Field, CA 94035, USA}
\author{Thayne Currie}  \affiliation{National Astronomical Observatory of Japan, Subaru Telescope, 650 A'ohoku Pl., Hilo, HI 96720, USA}
\author{Matthew A. Kenworthy}  \affiliation{Leiden Observatory, Leiden University, P.O. Box 9513, 2300 RA Leiden, The Netherlands}
\author{Christian Marois}  \affiliation{National Research Council of Canada Herzberg, 5071 West Saanich Rd, Victoria, BC, Canada V9E 2E7} \affiliation{University of Victoria, 3800 Finnerty Rd, Victoria, BC, Canada V8P 5C2}
\author{Johan Mazoyer} \affiliation{Johns Hopkins University, Zanvyl Krieger School of Arts and Sciences, Department of Physics and Astronomy, Bloomberg Center for Physics and Astronomy, 3400 North Charles Street, Baltimore, MD 21218, USA} \affiliation{Space Telescope Science Institute, 3700 San Martin Dr, Baltimore MD 21218, USA}
\author{Garreth Ruane} \affiliation{Department of Astrophysics, California Institute of Technology, 1200 E. California Blvd., Pasadena, CA 91101, USA}
\author{Angelle Tanner} \affiliation{Mississippi State University, Department of Physics \& Astronomy, Hilbun Hall, Starkville, MS, 39762, USA}

\begin{abstract}
As planning for the next generation of high contrast imaging instruments (e.g. WFIRST, HabEx, and LUVOIR, TMT-PFI, EELT-EPICS) matures, and second-generation ground-based extreme adaptive optics facilities (e.g. VLT-SPHERE, Gemini-GPI) are halfway through their principal surveys, it is imperative that the performance of different designs, post-processing algorithms, observing strategies, and survey results be compared in a consistent, statistically robust framework. In this paper, we argue that the current industry standard for such comparisons -- the contrast curve -- falls short of this mandate. We propose a new figure of merit, the ``performance map," that incorporates three fundamental concepts in signal detection theory: the true positive fraction (TPF), false positive fraction (FPF), and detection threshold.  By supplying a theoretical basis and recipe for generating the performance map, we hope to encourage the widespread adoption of this new metric across subfields in exoplanet imaging. 
\end{abstract}

\section{Introduction}
The contrast curve describes an imaging system's sensitivity for a given detection significance in terms of the planet/star flux ratio and angular separation. A consistent methodology for computing the contrast curve, however, is lacking: a variety of approaches to throughput, small sample-size, and non-Gaussian noise corrections are represented in the literature (e.g. \citealt{marois_confidence_2008}; \citealt{wahhaj_gemini_2013}; \citealt{mawet_fundamental_2014}; \citealt{pueyo_detection_2016}; \citealt{otten_-sky_2017}). As inner working angles are pushed below $5\lambda/D$, these details dominate the calculation of the contrast curve. Secondly, the contrast curve's information content is limited: by fixing the detection significance for all separations, the contrast curve conceals important trade offs between the choice of detection threshold, false positive rates, and detection completeness statistics.

The purpose of this paper is to critically examine the contrast curve and present alternative figures of merit for the ground and space-based exoplanet imaging missions of the coming decades. In Section \ref{sec:sdt}, we summarize the key points of signal detection theory, which provide the basis for our discussion of  performance metrics. Section \ref{sec:contrast} describes the strengths and weaknesses of the contrast curve as a general purpose performance metric. Finally, Sections \ref{sec:tradespace} and \ref{sec:PM_overview} give our proposal for a new figure of merit based on signal detection theory. 

\section{Overview of Signal Detection Theory}
\label{sec:sdt}
Our task as planet hunters is to decide whether the data at each location in a ``high contrast'' image meets our threshold for a planet detection. Regardless of the details of the dataset (e.g. field rotation, spectral coverage, etc.), the presence of noise will interfere with the accuracy of our detection decisions. Signal detection theory provides a precise framework for describing the relationships between detections, non-detections, and detection thresholds. 

If we assume that a planet is present at a location of interest in our data (the $H_1$, or ``signal present" hypothesis), and we succeed in detecting that planet, our result is a true positive (TP). If we fail to detect the planet, our result is a false negative (FN). Clearly, we aim to maximize the number of true positives while minimizing the number of false negatives. Hence, we define a true positive fraction, or TPF:
\begin{equation}
\mbox{TPF} = \frac{\mbox{TP}}{\mbox{TP} + \mbox{FN}} = \int^{+\infty}_{\tau} pr(x|H_1)dx
\end{equation}
where $\tau$ is the detection threshold and $pr(x|H_1)$ is the probability density function (PDF) of the data $x$ under the  hypothesis $H_1$. Our goal is to approach TPF$=1$.

\begin{figure}[h!]
    \centering
    \includegraphics[width=0.4\textwidth]{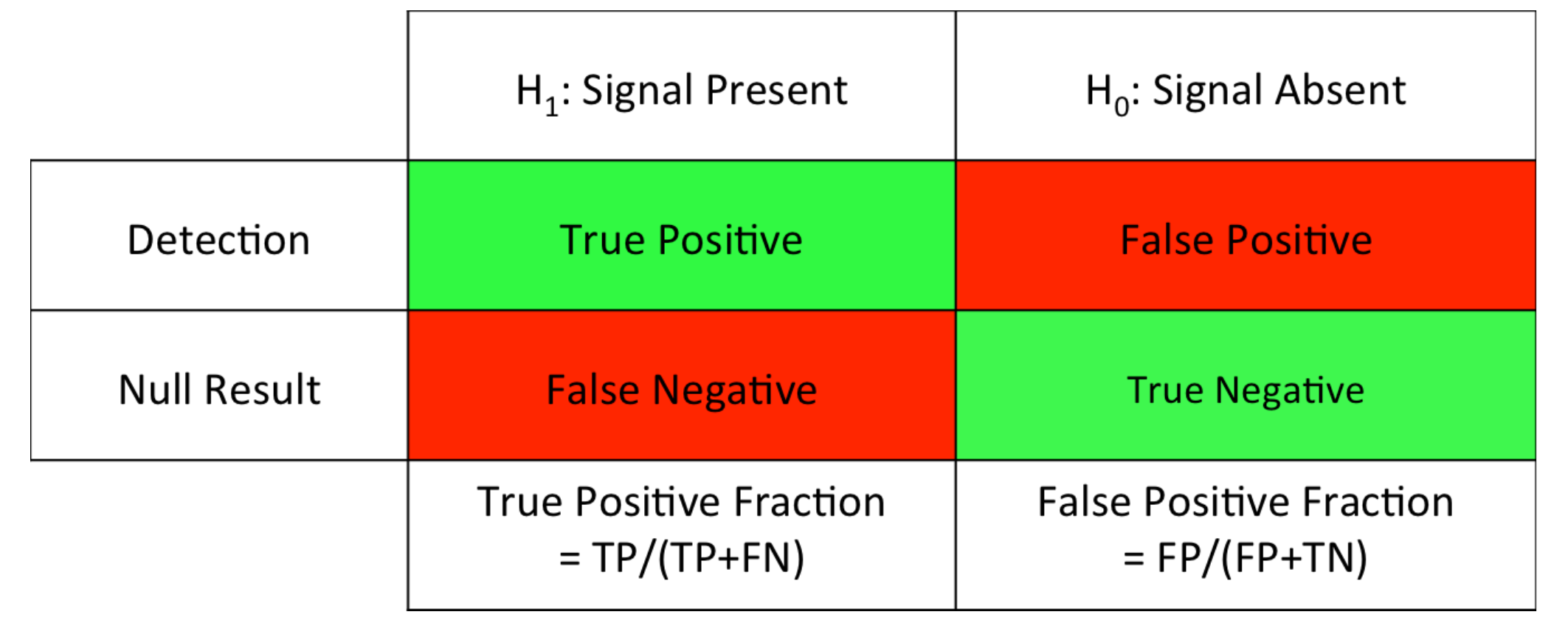}
    \caption{The confusion matrix}
    \label{fig:confusion}
\end{figure}

If we instead assume that no planet is present in the data (the $H_0$, or ``signal absent" hypothesis), and we fail to make a detection, our result is a true negative (TN). If we incorrectly claim to detect a planet, however, our result is a false positive (FP). We are then interested in achieving a false positive fraction (FPF) close to zero:
\begin{equation}
\mbox{FPF} = \frac{\mbox{FP}}{\mbox{TN} + \mbox{FP}} = \int^{+\infty}_{\tau} pr(x|H_0)dx.
\label{eqn:fpf}
\end{equation}
These various hypotheses and outcomes are summarized in the ``confusion matrix" (Figure \ref{fig:confusion}). An early review of signal detection theory is given by  \citet{swets_decision_1961}.

\begin{figure}[h]
  \centering
  \subfigure[Noise distribution]{\includegraphics[width=0.23\textwidth]{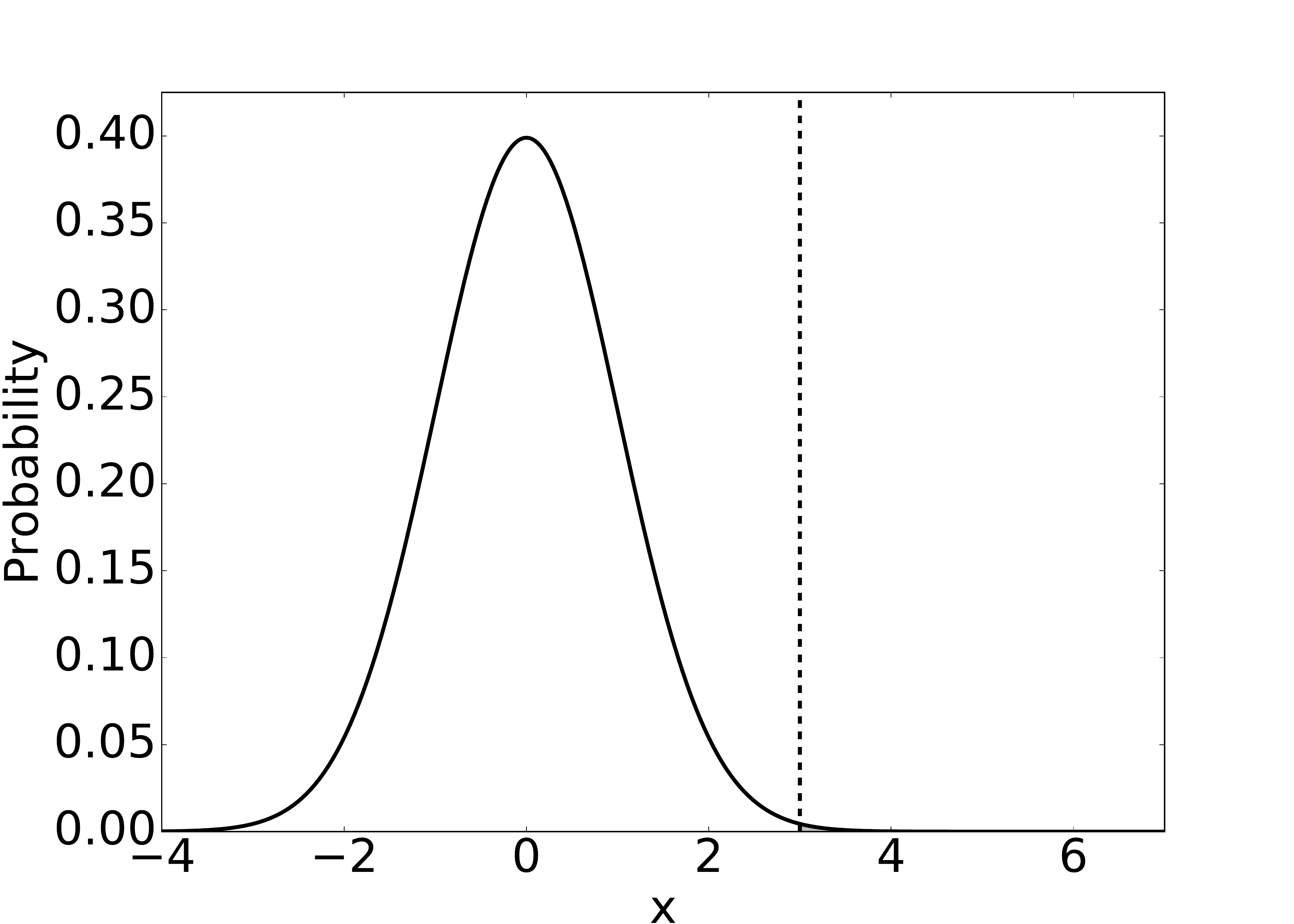}}
  \subfigure[Signal distribution]{\includegraphics[width=0.23\textwidth]{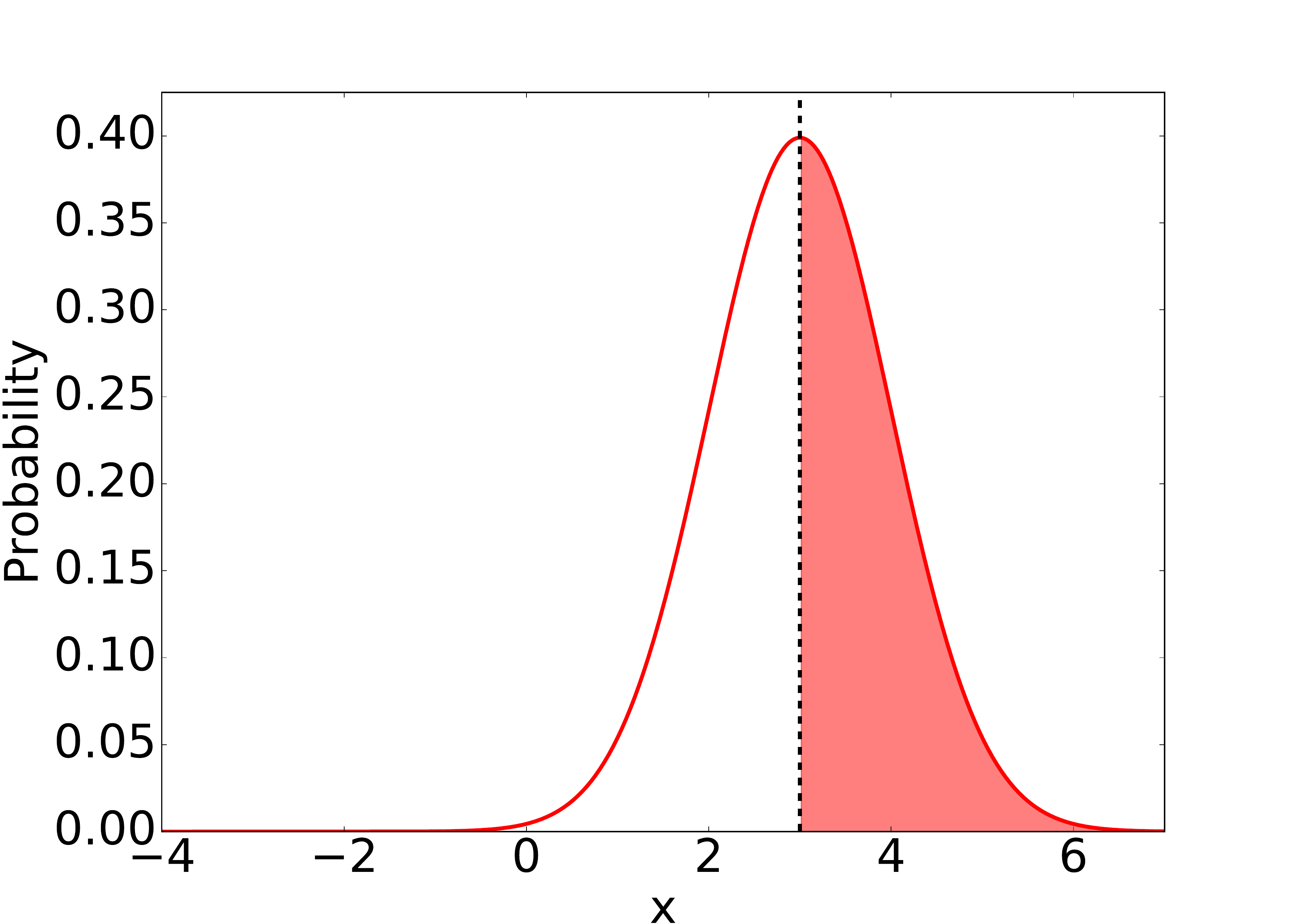}}
\caption{ (a) The normally distributed PDF of a noise source with a mean of zero and standard deviation of one. Here, the detection threshold is arbitrarily set to $3\sigma$ (dashed line), which corresponds to $x=3$ for this distribution. Because the noise PDF falls above the detection threshold a fraction $0.001$ of the time, the false positive fraction in this example is $0.001$. (b) The Gaussian PDF of a signal source with a mean of $x=3$ and a standard deviation of one. Because half of the signal distribution's area falls above the detection threshold, the true positive fraction is $0.5$. }
\label{fig:FPF_TPF}
\end{figure}

To make these relationships concrete, consider a post-processed image in which the intensities, $x$, in a series of photometric apertures located at a certain distance from the central star are drawn from a normal distribution ($\mu = 0$ and $\sigma = 1$, where the choice of an annular region is justified by the symmetry of the star's point spread function). The PDF of the noise is shown in Figure \ref{fig:FPF_TPF}a. Now let us assume that our goal is to detect a planet with a mean intensity of $x=3$ inside the annulus of interest. Because the intensity in the photometric aperture at the planet's location is also affected by the noise, it is described by a PDF identical to that of the noise, but with a mean of $x=3$ (here, we ignore the contribution of the planet's shot noise). The PDF of the signal is shown in Figure \ref{fig:FPF_TPF}b. 

Given our knowledge of the PDFs of the noise and the signal, we now wish to choose a detection threshold. Let us assume that because our detection follow-up resources (e.g. telescope time) are limited, we wish to achieve a false positive fraction of $0.001$. We therefore choose a detection threshold of $3\sigma$ because a fraction $0.001$ of the area of the noise PDF falls above this value (\ref{fig:FPF_TPF}a, dotted line). A second consequence of this choice of detection threshold is that we will only detect half of all planets with a mean intensity of $x=3$ (TPF$=0.5$; \ref{fig:FPF_TPF}b, dotted line). If we wish to increase the TPF, we must lower the detection threshold and hence unfavorably raise the FPF. 

\begin{figure}[h!]
    \centering
    \includegraphics[width=0.4\textwidth]{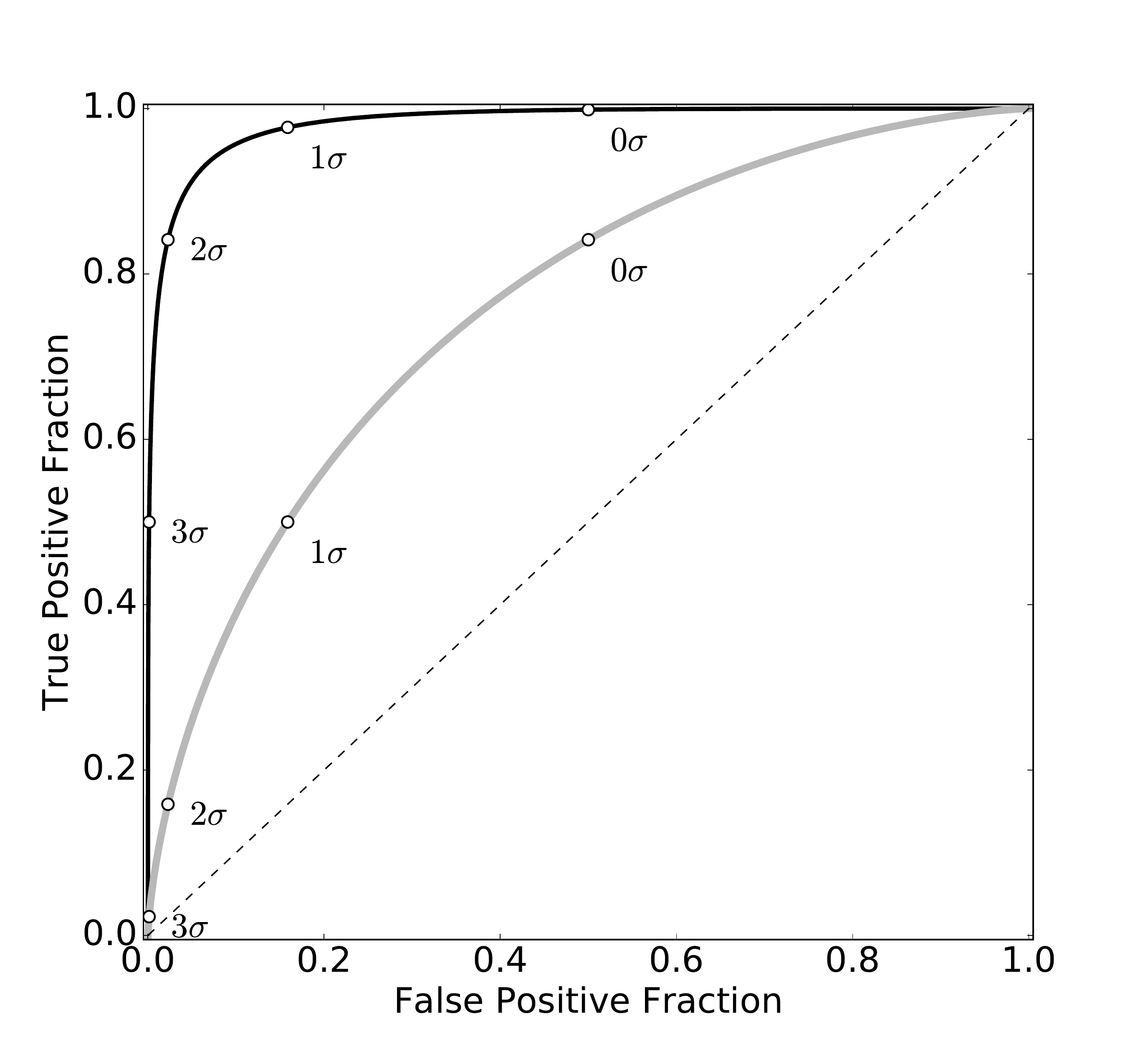}
    \caption{Black line: an ROC curve corresponding to a range of detection thresholds applied to the normal noise and signal distributions in Figure \ref{fig:FPF_TPF}. The (TPF, FPF) locations corresponding to thresholds of $0\sigma - 3\sigma$ are labeled to demonstrate the trade-offs between these key parameters. Grey line: the equivalent ROC curve for a signal distribution centered at $x=1$. }
    \label{fig:roc}
\end{figure}

Our choice of detection threshold therefore allows us to trade between the FPF and TPF, within the constraints imposed by the noise PDF and the signal mean. The receiver operator characteristic (ROC) curve allows us to visualize this trade by plotting the TPF as a function of the FPF, with each parameter varying between $0$ and $1$ as we move the detection threshold from large to small values (\citet{tanner_decision-making_1954} gives an early example of an ROC curve; \citet{krzanowski_roc_2009} provide an updated discussion of the topic). The black line in Figure \ref{fig:roc} shows the ROC curve associated with our example. The (TPF, FPF) pair corresponding to our example threshold of $3\sigma$ is labeled, along with a broader range of possible detection threshold choices. We note that the detection threshold must be less than the mean of the noise distribution to produce FPF values greater than 0.5. Because the mean is zero in this example in this example, such thresholds are negative. While mathematically consistent, negative thresholds have no observational relevance.

The shape of the ROC curve is determined by the shape of the noise distribution as well as the signal mean. For example, if we change the mean of the signal distribution in Figure \ref{fig:FPF_TPF}b from $x = 3$ to $x = 1$, we obtain the gray ROC curve shown in Figure \ref{fig:roc}. Because the noise distribution was unchanged, the black and gray curves' (TPF,FPF) pairs corresponding to detection thresholds of $0\sigma - 3\sigma$ share identical FPF values. Alternatively, if we had chosen a positively skewed rather than a normal noise distribution, the nearly vertical part of the black ROC curve at small FPFs would tilt to the right.

 We may now describe our goal of characterizing the detection statistics of an exoplanet imager in the vocabulary of signal detection theory: we wish to determine the maximum FPF and minimum TPF that satisfy our resource limitations and science goals -- in other words, we must choose a target location in (TPF, FPF) space. Our goal in designing an instrument, observing strategy, or post-processing routine is to produce a noise distribution whose ROC curve will reach that location for a signal of interest. 

An ROC curve, however, only represents a single noise distribution (i.e. image location) and signal level. In the sections that follow, we will discuss methods for representing the performance of a full image.

\section{Contrast Curves as Performance Metrics}
\label{sec:contrast}

\subsection{The Definition of the Contrast Curve}
The contrast curve is a means of representing the true and false positive fractions associated with a range of signals and positions in a final image. Schematically, we can define the contrast as:
\begin{equation}
\mbox{contrast} =  \left ( \frac{\mbox{factor} \times \mbox{noise}}{\mbox{stellar aperture photometry}} \right ) 
				   \left ( \frac{1}{\mbox{throughput}} \right )
\label{equ:contrast}
\end{equation}
where the numerator is the detection threshold, expressed as a multiple of the noise distribution's width. Often, the width of the noise distribution (here, the ``noise") is chosen to be the standard deviation of the resolution element intensities at a given separation from the star (e.g. Figure \ref{fig:rings}), while the multiplicative ``factor'' is chosen to be three or five to produce a $3\sigma$ or $5\sigma$ contrast curve. In Figure \ref{fig:FPF_TPF}, factor $=3$ and noise $=\sigma=1$. The detection threshold is then converted to a fraction of the parent star's brightness via the ``stellar aperture photometry" term. Finally, the ``throughput'' term corrects this brightness ratio for any attenuation of the off-axis signal relative to the star's (e.g. due to field-dependent flux losses imposed by the coronagraphic system and post-processing algorithms). The final contrast is therefore the planet-to-star flux ratio of a planet whose brightness is equal to the detection threshold. Figure \ref{fig:FPF_TPF} illustrates that the TPF associated with such a signal is 0.5. Hence, the contrast curve can be interpreted as the signal for which we achieve $50\%$ completeness given our choice of detection threshold in the numerator. The numerator also fixes the false positive fraction -- for example, choosing factor$=3$ for a white noise distribution gives FPF = 0.001. Finally, it is important to note that the contrast curve's statistics refer to planet detectability, and not to the photometric accuracy associated with any given planetary signal.

\subsection{Where the Contrast Curve Falls Short}
\label{sec:contrast_problems}

Both practical and fundamental shortcomings, however, undermine the utility of the contrast curve as a general purpose performance metric. First, the contrast is inflexible: by fixing the true positive fraction to 0.5 and the false positive fraction to a value set by the numerator, we cannot explore the (TPF, FPF, detection threshold) trade space. Even if we were to plot multiple contrast curves on the same figure to show different detection thresholds, we could not escape the arbitrary choice of TPF$=0.5$. Similarly, if we were to plot a $90\%$ detection completeness curve as a function of separation, we could not access a range of false positives fractions. Finally, fixing the TPF, FPF, and detection threshold for all separations may not be desirable for all applications -- because the number of resolution elements, the PDF of the noise, and the predicted population of planets all vary as a function of separation, a particular imaging program's science goals may be better served by a detection threshold that also varies with separation.

More problematic, however, is the calculation of the terms in Equation \ref{equ:contrast}. As mentioned above, the ``noise'' term is typically chosen to be the standard deviation of resolution elements in a region of the image, whose shape and size widely varies in the literature. This approach is valid if two conditions are met: 1) if the region includes enough statistically independent realizations of the noise to allow for an accurate measure of the distribution's standard deviation, and 2) if the underlying noise distribution is Gaussian. While there is no hard and fast rule for deciding whether the first condition is met, statisticians generally consider 30 independent samples to be the boundary between large and small sample statistics \citep{wilcox_basic_2009}. For the case of $1\, \lambda/$D-wide annular regions, 30 samples corresponds to a separation of $\sim 5 \lambda/$D. Below this threshold, the sample standard deviation is an increasingly uncertain estimate of the width of the underlying noise distribution \citep{student_probable_1908,mawet_fundamental_2014}. The mitigating strategy proposed by \citet{mawet_fundamental_2014}, however, also requires that condition $\#2$ (Gaussian noise) is met. \citet{aime_usefulness_2004} and many others have shown that uncorrected low-order wavefront aberrations cause the noise at small separations to follow a positively skewed modified rician distribution rather than a normal distribution \citep{perrin_structure_2003, bloemhof_anomalous_2004, fitzgerald_speckle_2006, soummer_speckle_2007, hinkley_temporal_2007,marois_confidence_2008}. While numerous observing and post-processing strategies have been employed to whiten this skewed distribution (e.g. \citealp{liu_substructure_2004}; \citealp{marois_angular_2006}; \citealp{lafreniere_new_2007}; \citealp{amara_pynpoint:_2012}; \citealp{soummer_detection_2012}), their success at small separations is limited by the temporal and spectral variability of the noise (Appendix \ref{app:SW} discusses the difficulty of testing for normality using methods such as the Shapiro-Wilk test). The result is that the noise distribution at small angles retains an unknown skewness at small separations that increases the false positive fraction compared to a Gaussian distribution.  Hence, neither condition for the use of the standard deviation as a proxy for the FPF is met at small separations\footnote{It is worth noting that some authors interpret the numerator of Equation \ref{equ:contrast} as an empirical signal to noise threshold without reference to the distribution of the noise or a false positive fraction. This interpretation, however, robs the contrast curve of much of its practical use -- the knowledge that we can achieve TPF$=0.5$ for a given planet:star flux ratio does not guide our observing or science if the associated false positive fraction can fall anywhere from zero to one.}. In Section \ref{sec:smallseps} we will address alternative methods for probing the distribution of the noise without the assumption of normality. 

\subsection{Inconsistencies in Contrast Curve Computations}

We further note that the contrast and its constituent terms are inconsistently computed in the literature, in particular the noise and throughput terms. While many authors (e.g. \citealp{wahhaj_gemini_2013}) account for spatially correlated speckle statistics by defining the noise to be the standard deviation of resolution elements in an annulus, others do not. For example, \citet{otten_-sky_2017} define the noise in relation to the standard deviation of pixel values inside of a single $1\, \lambda/$D aperture of interest. The region within a few $\lambda/$D of the inner working angle, however, is fundamentally  sensitive to azimuthally correlated speckle noise: effects such as pointing jitter, thermal variations, and non-common path aberrations induce low order wavefront aberrations, and hence close-in, variable speckles, on the timescale of an observation \citep{shi_low-order_2016}.  Secondly, the definition of the term ``throughput" is context dependent. Authors computing contrast curves for angular differential imaging (ADI) datasets typically define the throughput in terms of the flux losses imposed by signal self-subtraction (e.g. \citealp{wahhaj_gemini_2013}). However, in discussions of coronagraph design trades, throughput refers to the often field-dependent flux losses caused by the coronagraphic system itself (e.g. \citealp{guyon_theoretical_2006}; \citealp{krist_numerical_2015}). Finally, the small sample correction presented by \citet{mawet_fundamental_2014} has been adopted by some authors (e.g. \citealp{wertz_vlt/sphere_2016}), but not others (e.g. \citealp{uyama_seeds_2016}). Such a variety of methodologies inhibit meaningful comparisons of instrument performance. 

In this section, we have described three shortcomings of the contrast curve: 1) its inability to illustrate the (TPF, FPF, detection threshold) trade space, 2) its potential inconsistency with the shape of the underlying noise distribution, and 3) its inconsistent treatment in the literature. In the sections that follow, we will discuss strategies for computing the FPFs and TPFs associated with an unknown noise distribution and present a new figure of merit for the performance of high dynamic range imaging systems.  

\subsection{The Raw Contrast}
The above discussions concern what we might call an ``observer's" definition of the contrast. Users of exoplanet imaging testbeds, however, refer to the ``raw contrast,'' which is typically defined as 
\begin{equation}
\mbox{raw contrast} = \frac{\mbox{mean}[R(x,y)]}{\mbox{max}[\mbox{PSF}_{\mbox{star}}(x,y)]}
\end{equation}
where $\mbox{mean}[R(x,y)]$ is the mean number of photons per pixel over a region of interest (for example a dark hole) and $\mbox{max}[\mbox{PSF}_{\mbox{star}}(x,y)]$ is the number of photons in the pixel corresponding to the peak of a stellar PSF offset to a representative location inside of the region of interest. The key difference between the raw contrast and the observer's contrast is that the raw contrast does not refer to an astrophysical flux ratio -- a raw contrast of $10^{-10}$ does not indicate that a planet with an astrophysical flux ratio of $10^{-10}$ is in any sense detectable. Rather, it simply indicates that the mean intensity of the background in a certain region is $10^{10}$ smaller than the peak of the offset stellar PSF. Hence, while the raw contrast is a useful shorthand for describing an instrument's starlight suppression, it should not be interpreted as a detection limit. Obtaining a detection limit by estimating the noise inside of the region of interest carries with it the attendant dangers of small sample statistics and non-Gaussian noise described in Section \ref{sec:contrast_problems} as well as exposure time dependencies and signal throughput effects.

\section{Representing the (FPF, TPF, separation) Tradespace with the Performance Map}
\label{sec:tradespace}
In Section \ref{sec:contrast_problems}, we argued that the contrast curve's limited information content -- the astrophysical flux ratios of those planets that give TPF=0.5 for a single detection threshold as a function of separation -- obscures the much richer (FPF, TPF, separation) trade-space. Here, we propose two modifications to the contrast curve: 1) a detection threshold (and hence FPF) that varies with separation, and 2) the inclusion of all possible TPFs as a heatmap.

When the detection threshold is held constant with separation, the radial distribution of false positives is not uniform because the number of resolution elements varies with separation. If the expected number of false positives $N_{\text{FP}}$ is given by $N_{\text{FP}}(r) = \text{FPF} \times 2 \pi r$ for separation $r$, then a constant detection threshold (and hence a constant FPF) allows more total false positives at wide separations than at small separations. If we instead keep the radial distribution of false positives constant, we allow the detection threshold to adapt to the changing number of resolution elements with separation (see \citet{2017arXiv170607489R} for a similar approach). 

Next, we plot the astrophysical flux ratios of those planets that give any desired TPF as a function of separation. Rather than choosing a single TPF contour, we propose to show the full $0 \leq \text{TFP} \leq 1$ space as a heatmap. A representative TPF contour can be overplotted for clarity. 

We call this modified figure the performance map (e.g. Figure \ref{fig:gaussian_PM_0.1fps}). We argue that the performance map highlights the most scientifically and programmatically relevant quantities, namely the TPFs of the signals of interest for a given number of false positives. The contrast curve, on the other hand, highlights the detection threshold, which has no intrinsic meaning beyond pointing to a false positive fraction. 

\section{Generating the Performance Map}
\label{sec:PM_overview}

Constructing the performance map requires knowledge of the false positive fraction which in turn requires knowledge of the underlying noise distribution. As discussed in Section \ref{sec:contrast_problems}, the distribution of the noise at small separations is often unknown. In the following subsections, we consider two limiting cases: 1) the PDF of the noise is Gaussian (Section \ref{sec:assumegaussian}), and 2) the PDF of the noise is completely unknown (Section \ref{sec:smallseps}). 

\begin{figure}[h]
    \centering
    \includegraphics[width=0.3\textwidth]{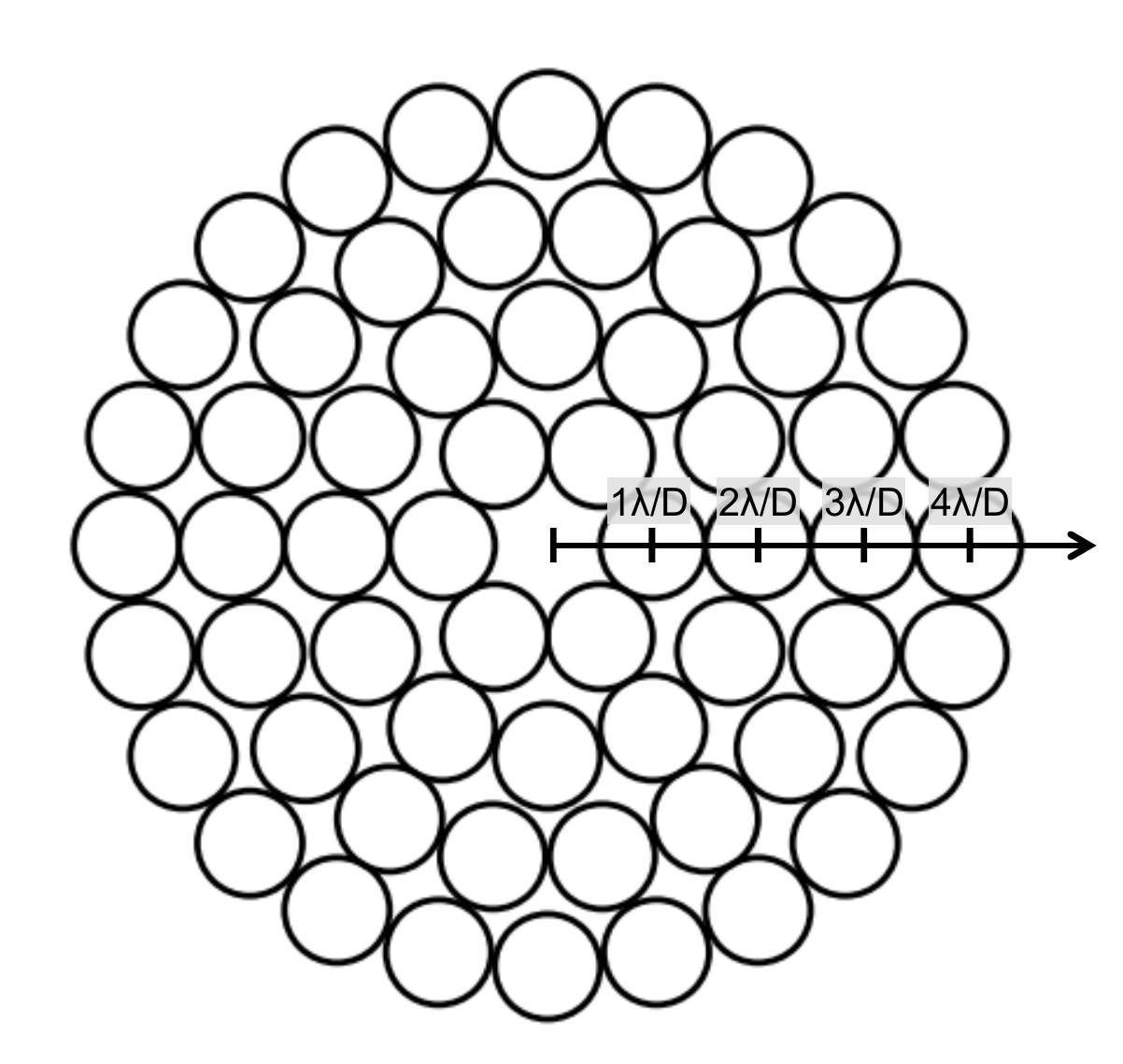}
    \caption{The number of resolution elements of width $\lambda/$D at a distance $r$ from the central star is $2 \pi r$, where here we consider only whole numbers of resolution elements.}
    \label{fig:rings}
\end{figure}

Following \citet{mawet_fundamental_2014}, we define a resolution element to be a circular aperture with a diameter of $\lambda/$D. The number of resolution elements, $N_r$ at a distance $r$ from the central star is $2\pi r$, where $r$ is also expressed in terms of $\lambda/$D (Figure \ref{fig:rings}). We consider only whole numbers of resolution elements (e.g. six resolution elements at $1\lambda/$D.). 

To illustrate the construction of a performance map in detail, we consider a set of HR8799 observations taken by the Spectro-Polarimetric High-contrast Exoplanet REsearch (SPHERE, \citealp{beuzit_sphere:_2008}) at the Very Large Telescope (VLT). The data were acquired in December of 2014 during science verification of the Infra-Red Dual-band Imager and Spectrograph (IRDIS, \citealp{dohlen_infra-red_2008}) instrument, and have been extensively described in the literature \citep{zurlo_first_2016,apai_high-cadence_2016,wertz_vlt/sphere_2016}. We adopt a 200 frame broadband H filter ($1.48-1.77\,\mu$m) sequence from December 4th 2014, where the detector integration time was $8\,$s and the total amount of parallactic angle rotation was $8^{\circ}.7$. We choose to include only the data taken on the left-hand side of the IRDIS detector. 

Following \citet{wertz_vlt/sphere_2016}, we use an off-axis broadband H image of $\beta$ Pictoris (January 30th 2015, PI: A.-M. Lagrange) as our PSF template due to the absence of an off-axis exposure in the original observing sequence. We fit a 2D Gaussian function to the $\beta$ Pictoris template PSF to obtain FWHM $=4.0\,$pixels $=0^{\prime\prime}.049$ for a plate scale of $12.251\,$mas \citep{wertz_vlt/sphere_2016}. Because this measured FWHM is slightly larger than the diffraction limit would suggest ($0.98 \times \lambda/$D$=0^{\prime\prime}.040$), we conservatively adopt the FWHM as the resolution element diameter rather than $1\, \lambda/$D.

For the purposes of this demonstration, we are interested in estimating the FPFs and planet-injected TPFs. Hence, we begin our reduction by subtracting HR$8799\,$bcde from the dataset. This is accomplished via the Vortex Image Processing (VIP, \citealp{2017AJ....154....7G}) package's functions for injecting negative fake companions into the data and optimizing their flux and positions using a Nelder-Mead based minimization. 

Next, we use VIP's implementation of the PCA-ADI algorithm to subtract a reconstructed datacube from our set of 200 images. The reconstructed cube was generated using three principal components (chosen to maximize the SNR of HR$8799\,$c in a full reduction of the dataset prior to planet subtraction). We median-combine the residual datacube to obtain our final reduced image. We compute the algorithmic throughput (signal self-subtraction) as a function of separation by injecting fake planets at separation intervals of $1\,$FWHM and azimuthal intervals of $120^{\circ}$. For each separation interval, the data is PCA-ADI reduced, and the signals' flux attenuation in the three azimuthally separated apertures are averaged.

\begin{figure}[h]
    \centering
    \includegraphics[width=0.45\textwidth]{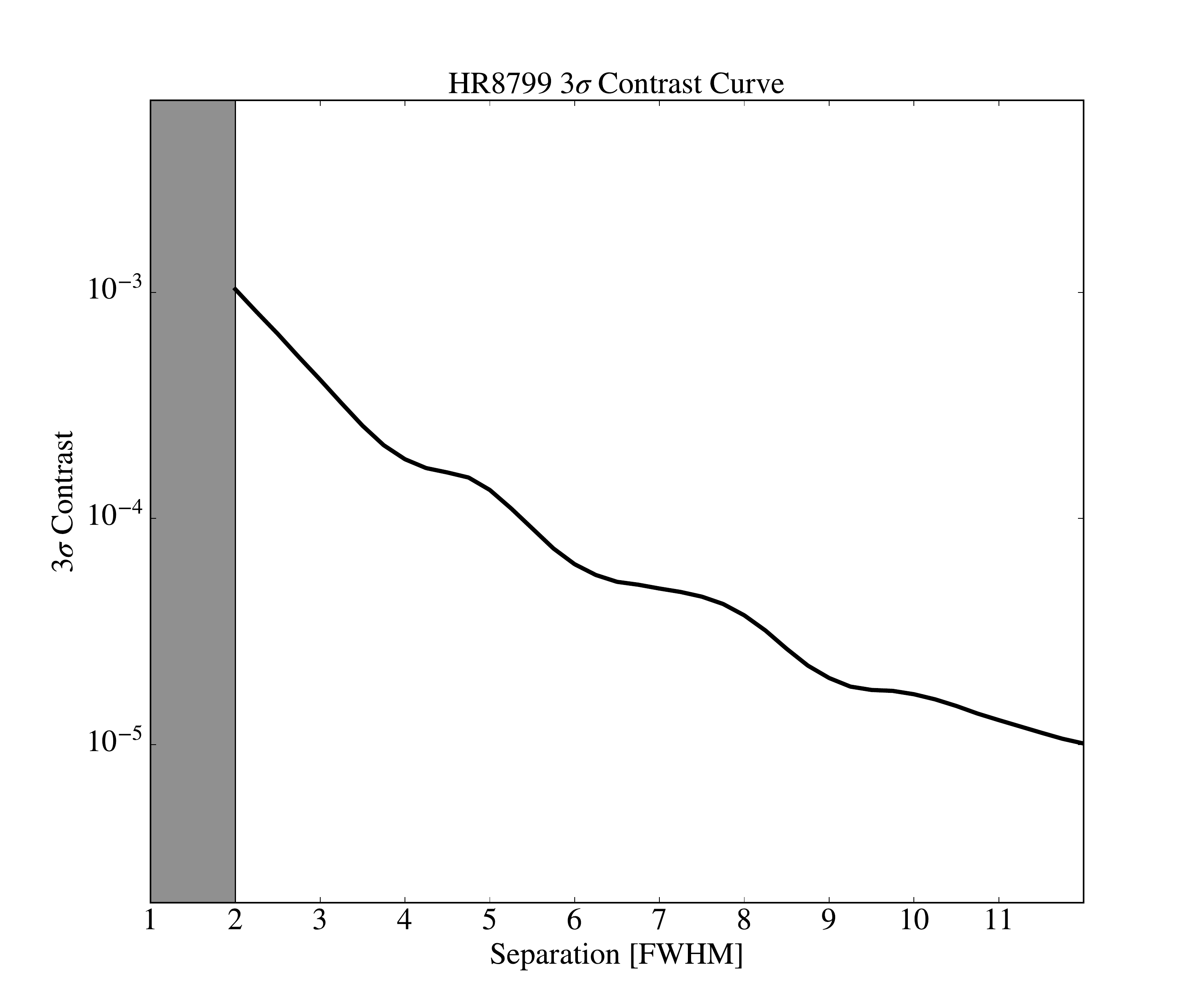}
    \caption{A contrast curve representing the observation of HR8799 with SPHERE described in Section \ref{sec:PM_overview}.}
    \label{fig:hr8799_contrast}
\end{figure}

Here, we consider only the first ten separation intervals after the inner working angle (in this case FWHM$=2-11$). Figure \ref{fig:hr8799_contrast} shows a $3\sigma$ contrast curve generated with VIP (where algorithmic throughput and small sample statistics are properly accounted for). 

\subsection{The Gaussian Assumption}
\label{sec:assumegaussian}

We first consider the most straightforward path to constructing the performance map: assuming that the PDF of the noise follows a Gaussian distribution with a width corresponding to the measured standard deviation of the resolution elements as a function of separation (acknowledging the uncertainty in the standard deviation due to small sample statistics). Because any calculation of the FPF requires the hypothesis $H_0$ (signal absent), we are assuming that any detections are false, despite the reality that there may be true planets in the data.

To choose the FPF (and hence the detection threshold) for each separation, we must first choose the total number of false positives that we are willing to accept in the FWHM$=2-11$ region of interest. For example, perhaps we have sufficient telescope time to follow up one false positive for every ten observations. Hence, we can accept 0.1 false positives per image, or FPF$ = \mathbf{0.01}/N_r$.  For each separation we then derive the corresponding detection threshold that will connect the FPFs to the TPFs of the injected signals. Here, the threshold is given by the quantile (inverse CDF) function of the Student T distribution with $N_r$ degrees of freedom and a width scaled by the measured standard deviation of the resolution elements at $r$. We can then inject planet signals to determine the TPF of a given signal at a given separation. For the purposes of this simplified demonstration, the TPF is computed using the CDF of the scaled Student T distribution representing the noise, but shifted by the throughput-corrected test signal. 

\begin{figure}[h]
    \centering
    \includegraphics[width=0.5\textwidth]{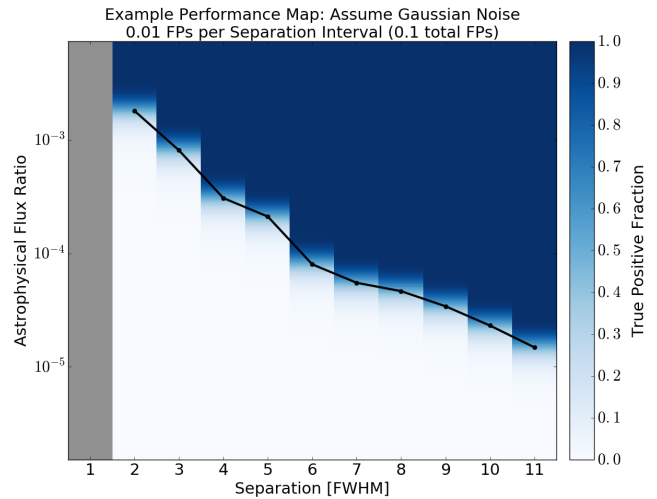}
    \caption{An example performance map where the FPFs have been calculated under the assumption that the noise is Gaussian at all separations. Here, the detection thresholds that connect the FPFs to the TPFs of the injected signals were chosen to give 0.01 false positives per separation interval, or 0.1 total false positives in the FWHM$=2-11$ region. }
    \label{fig:gaussian_PM_0.1fps}
\end{figure}

The resulting performance map is shown in Figure \ref{fig:gaussian_PM_0.1fps} with the TPF$=0.5$ contour overplotted. 

We emphasize that the ``depth" of the TPF$=0.5$ contour in Figure \ref{fig:gaussian_PM_0.1fps} is different from that of the contrast curve in Figure \ref{fig:hr8799_contrast} because the performance map is illustrating a lower false positive rate in this example. Furthermore, the performance map allows the detection threshold to vary with separation, while the contrast curve holds the detection threshold fixed.

\subsection{The Empirical Performance Map}
\label{sec:smallseps}

In the preceding section, we considered an ideal scenario in which the PDF of the noise was known and the false positive fractions could be computed analytically. In Section \ref{sec:contrast_problems}, however, we argued that the PDF of the noise at small separations is difficult to determine given the effects of imperfect speckle subtraction. 

To address this effect, we now consider an extreme case where the PDF of the noise is completely unknown, and the FPFs must be determined empirically: for each separation, we will simply count the number of resolution elements that exceed a test detection threshold. For a single $1\, \lambda/$D-thick annulus in the final, post-processed image, the possible values of the empirical false positive fraction are therefore constrained to $i/N_r$, where $i$ is an integer between zero and $N_r$ (inclusive). Accessing desirable FPFs between zero and $1/N_r$ requires additional realizations of the noise -- for example, data from the same instrument can provide additional resolution elements if the distribution of the noise is assumed to be constant with time. \citet{2017ApJ...842...14R} describe the application of this technique to the Gemini Planet Imager Extra Solar Survey (GPIES) campaign. Another possibility for the case of ADI data is obtaining an additional image ``for free" by reversing the order of the parallactic angle assignments \citep{wahhaj_gemini_2013}. This produces an image with similar azimuthal noise characteristics to the science image, doubling the number of noise realizations. Further angle randomization, however, will artificially whiten the speckle noise and fail to capture the temporal speckle evolution that de-rotation translates into azimuthal variation. 

To generate a performance map from a single image using this empirical FPF approach, we first make a list of FPFs for a range of detection thresholds and separations by the following recipe:
\begin{enumerate}
\item Draw rings of FWHM-diameter apertures around the central star (see Figures \ref{fig:rings}) and sum the fluxes inside of the apertures. The result is a list of $2 \pi r$ aperture sums for each separation $r$.
\item Choose a detection threshold.
\item For each separation, find the fraction of resolution elements whose sum exceeds the detection threshold. These are the FPFs.
\item Vary the detection threshold and repeat Step 3 to produce all possible FPF values for all separations. 
\end{enumerate}
Using the same set of detection thresholds and separations as the preceding recipe, we can find the associated TPFs for a range of planet signals of interest. This is accomplished by the following steps: 
\begin{enumerate}
\item Sum the flux inside of a FWHM-diameter aperture around the unocculted stellar PSF\footnote{As mentioned above, our example dataset lacks an unocculted image, but we fit the positions and fluxes of the HR$8799$ planets using a later off-axis observation of $\beta$ Pictoris. For the purposes of this example, we estimate HR$8799$'s unocculted aperture sum based on the fitted flux of HR$8799\,$b and the H-band planet-to-star flux ratio given in \citet{marois_direct_2008}.}. 
\item Choose an astrophysical flux ratio and multiply by the star's aperture sum (previous step) to obtain the planet's signal.
\item For each separation, multiply the planet's signal by the algorithmic throughout (previous paragraph), and add the result to each resolution element. 
\item Choose a detection threshold from the same list of threshold used to generate the FPFs above.
\item For each separation, find the fraction of resolution elements whose sum exceeds the threshold. These are the TPFs. 
\item For the same range of detection thresholds used to calculate the FPFs, repeat Step 5.
\item Repeat Steps 2-6 for different astrophysical flux ratios. 
\end{enumerate}

\begin{figure}[h]
    \centering
    \includegraphics[width=0.5\textwidth]{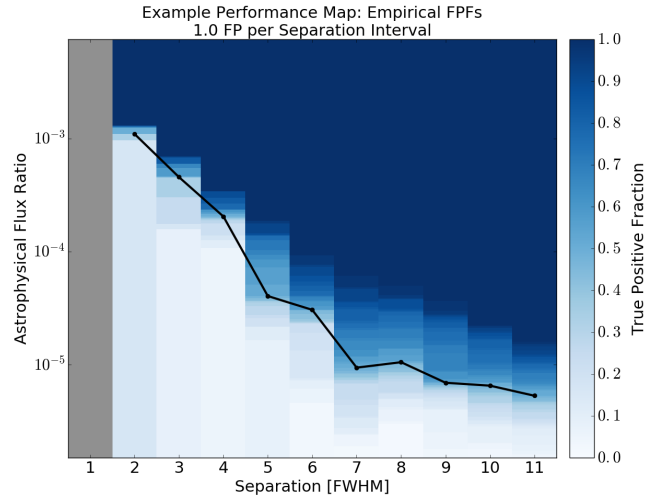}
    \caption{The performance map shows the astrophysical flux ratio versus the separation, color-coded by the true positive fraction. The solid black line represents the approximate TPF $=0.5$ contour.}
    \label{fig:perf_map}
\end{figure}

To plot the performance map, we elect to consider the smallest detection thresholds associated with the least non-zero FPFs ($1/N_r$), giving 1.0 false positive per $1\, \lambda/$D separation interval. For each injected signal at each separation, we then plot the TPFs corresponding to these detection thresholds. Figure \ref{fig:perf_map} shows the resulting performance map. For each separation, we also plot the signal with the TPF nearest to TPF$=0.5$ for these detection thresholds.

We may now compare the total number of false positives in the empirical performance map above with that of the contrast curve. The choice the $3\sigma$ detection threshold used to compute the contrast curve implies a false positive fraction of 0.0013 under the assumption that the noise is Gaussian. To obtain the total number of false positives in the image under this assumption, the false positive fraction is multiplied by the total number of whole resolution elements, $N_T$. For the $2-11\,\lambda/$D region of the image considered here, $N_T=403$ and the total number of false positives is $N_{FP} = FPF \times N_T \approx 0.54$. In comparison, the empirical performance map approach makes no assumptions about the PDF of the noise, and gives one false positive per separation, or $N_{FP}=10$ in this example. Given additional realizations of the noise (e.g. observations of other targets in a homogeneous survey), however, the least non-zero FPF is $1/(N_rN_f)$, where $N_f$ is the total number of frames. In this example, reducing $N_{FP}$ from 10 to the $3\sigma$ white noise case of $N_{FP}=0.54$ would require $10/0.54 \approx 19$ additional images to increase the total number of resolution elements available at each separation. 

While the empirical approach described here does require many observations to reach the small FPFs promised by the Gaussian noise assumption, it will eventually yield the correct connections between the FPFs, detection thresholds, and TPFs as the number of images increases, regardless of the underlying PDF of the noise. Hence, such an approach is particularly appealing for large surveys, and less appealing for a single observation.

\section{Hypothesis Testing}
In the discussion above, our calculation of the true and false positive fractions required a choice of hypothesis: either $H_1$ (signal present; planet-injected data), or $H_0$ (signal absent; planet-free data). This approach allowed us to characterize the performance of the instrument by probing the range of possible TPFs and FPFs for various positions and signals.

We may also consider a different objective: deciding whether a particular bright spot in our final science image is a planet. In this scenario, we must decide which hypothesis, $H_1$ or $H_0$, applies to our location of interest. While hypothesis testing is beyond the scope of this paper, we refer to the detailed discussions in \citet{kasdin_linear_2006}, Section 5, and \citet{young_image_2013}.

\section{Conclusion}
As the cost and complexity of ground and space-based exoplanet imaging missions increase, so too must the fidelity and relevance of our diagnostic tools improve. We argue that the drawbacks of the contrast curve -- its lack of transparency, flexibility, and connection to the data -- motivate a re-evaluation of its use as a general purpose performance metric. Our proposed ``performance map'' is one among many possible methods for visualizing the true and false positive fractions associated with a high dynamic range image. The performance map is an opportunity for displaying the results of planet search programs in a consistent and statistically correct way as well as comparing the performance of various post-processing algorithms within a well-defined statistical framework. By encouraging the scrutiny of this new metric, we hope to improve the prediction and evaluation of the performance of the next generation of high contrast imaging instruments.

\acknowledgments
This work has been supported by the Keck Institute for Space Studies and the NASA Exoplanet Exploration Program Study Analysis Group $\#19$: Exoplanet Imaging Signal Detection Theory and Rigorous Contrast Metrics. This material is based upon work supported by the National Science Foundation Graduate Research Fellowship under grant No. DGE-1144469. G. Ruane is supported by an NSF Astronomy and Astrophysics Postdoctoral Fellowship under award AST-1602444. C.G.G. and O.A. ackowledge funding from the European Research Council Under the European Union's Seventh Framework Program (ERC Grant Agreement n.\ 337569) and from the French Community of Belgium through an ARC grant for Concerted Research Action.

\software{Vortex Image Processing \citep{2017AJ....154....7G}}


\appendix
\section{The Shapiro-Wilk Test}
\label{app:SW}

\begin{figure*}[t]
  \centering
  \subfigure[The PDF of a normal distribution (black line) and Rayleigh distribution (red line) with a scale parameter of two.]{\includegraphics[width=0.45\textwidth]{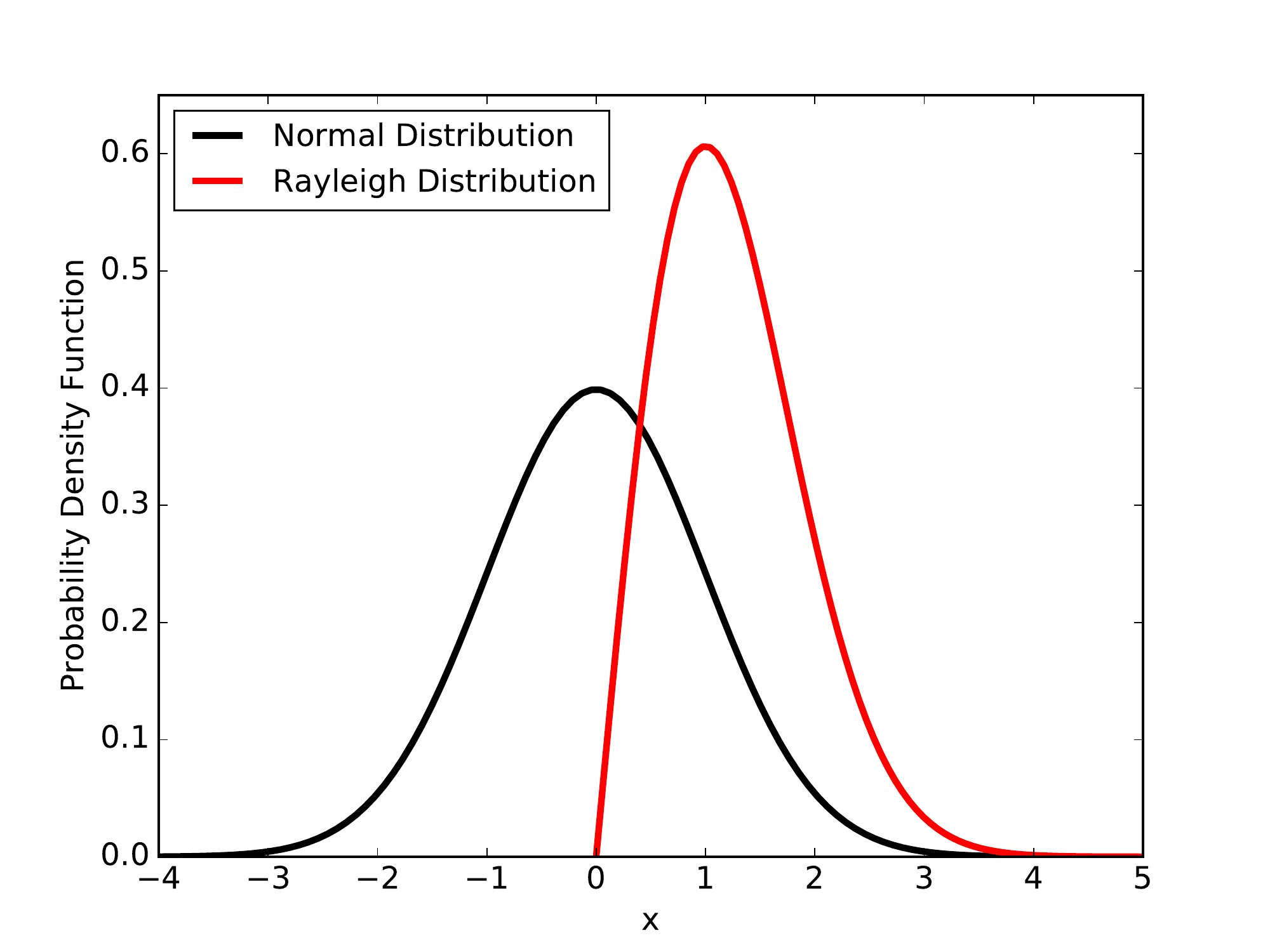}}
  \subfigure[The fraction of all trials that reject the null hypothesis ($p \leq 0.01$) for the normally distributed data (black circles) and Rayleigh distributed data (red stars). The black dotted line indicates the expected fraction of trails for which the normally distributed data is expected to reject the null hypothesis ($p=0.01$).]{\includegraphics[width=0.45\textwidth]{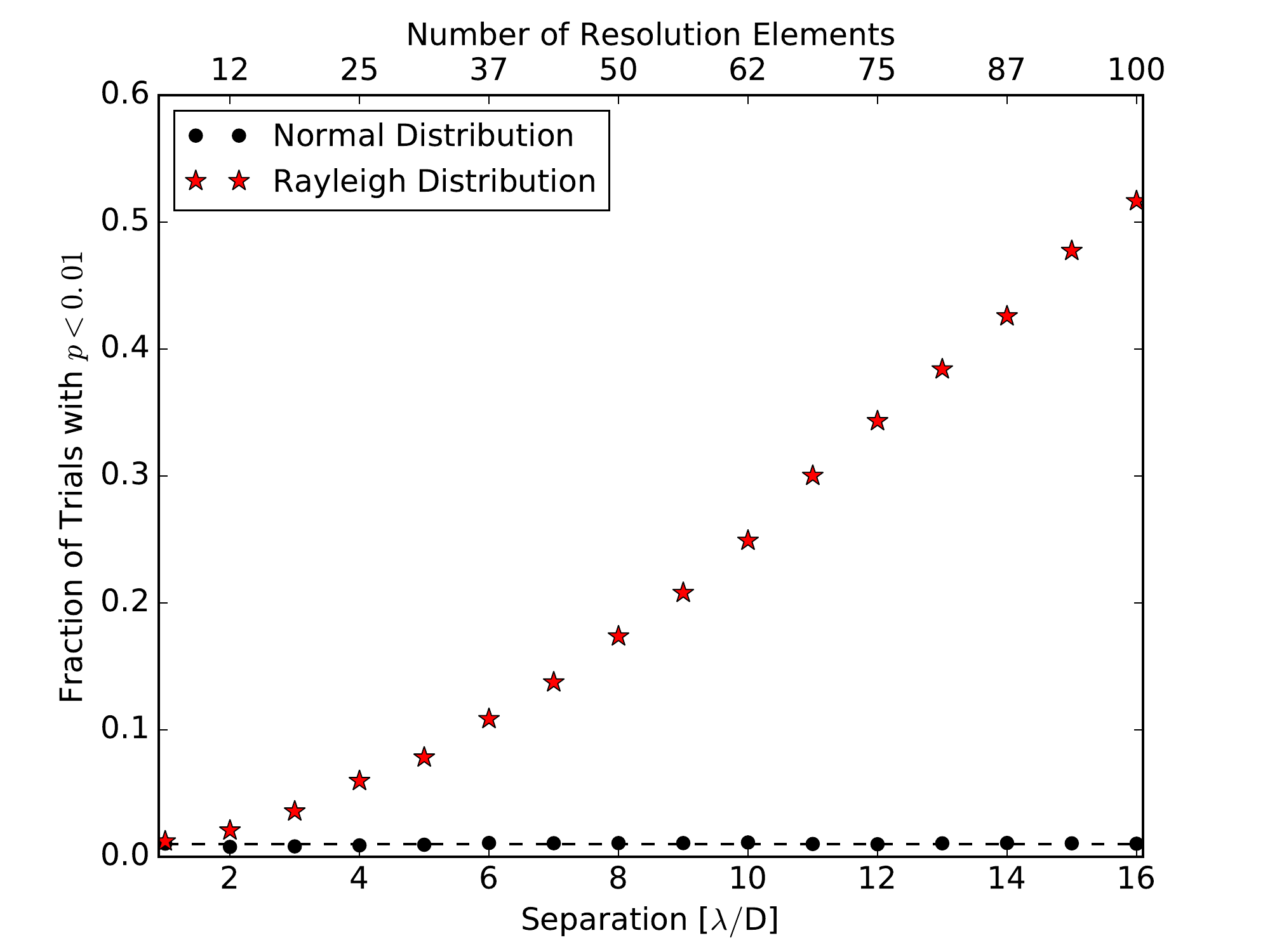}}
\caption{Even though the Rayleigh distribution (scale parameter = 2) is highly skewed compared with the normal distribution, the Shapiro-Wilk test cannot reliably distinguish it from a normal distribution for the sample sizes shown here. For separations less than $15\, \lambda/$D, the Shapiro-Wilk test gives the wrong outcome (fails to reject the null hypothesis) for more than half of all trials.}
\label{fig:SW}
\end{figure*}

For a given post-processed dataset, we may be interested in testing whether our data has been successfully whitened at small separations. The Shapiro-Wilk test \citep{shapiro_analysis_1965} tests the null hypothesis that a dataset was drawn from a normal distribution; it returns a $p-$value that specifies that probability of obtaining the dataset given the null hypothesis. In order to test the utility of the Shapiro-Wilk test at small separations, we consider data drawn from two different distributions: a normal distribution (Figure \ref{fig:SW}a, black line), and a positively skewed Rayleigh distribution (Figure \ref{fig:SW}a, red line). At face value, we expect to easily reject the Shapiro-Wilk test's null hypothesis when testing data drawn from the dramatically non-white Rayleigh distribution.

We first compute the Shapiro-Wilk test $p-$value using a normally distributed dataset with $2 \pi r$ elements. We then draw new sets of $2 \pi r$ elements to repeat the test $10^4$ times, giving $10^4$ $p-$values per separation. We arbitrarily choose $p \leq 0.01$ as our threshold for rejecting the null hypothesis. As expected, we find that for all separations, the normally distributed test data gives $p \leq 0.01$ a fraction $0.01$ of the time (Figure \ref{fig:SW}b, black points). 

Next, we repeat this procedure for the Rayleigh distributed data. We find that these data reject the null hypothesis for a much larger fraction of trials than the normally distributed data (Figure \ref{fig:SW}b, red points). However, we quickly see a problem: at $15\lambda/$D, the Rayleigh distributed data only rejects the null hypothesis about half of the time. This means that in any one science image, the probability of erroneously accepting the null hypothesis that the data is normally distributed is also $50\%$. At smaller separations, we draw the wrong conclusion most of the time -- hence, the Shapiro-Wilk test cannot be used to test for normality at small separations.

Some tests (e.g. the Kolmogorov-Smirnov test) perform better in these respects than the Shapiro-Wilk test, while others (e.g. the Anderson-Darling test) are similarly problematic. The purpose of the example given here is to demonstrate that the outcomes of normality tests cannot be taken at face value, and must be rigorously validated in order to be applied to  observational datasets. 

\end{document}